\input harvmac.tex

\lref\DixonGR{
  L.~J.~Dixon, L.~Magnea and G.~Sterman,
  ``Universal structure of subleading infrared poles in gauge theory
  arXiv:0805.3515 [hep-ph].
}

\lref\MochPA{
  S.~Moch, J.~A.~M.~Vermaseren and A.~Vogt,
  ``The three-loop splitting functions in QCD: The non-singlet case,''
  Nucl.\ Phys.\  B {\bf 688}, 101 (2004)
  [arXiv:hep-ph/0403192].
}

\lref\VogtMW{
  A.~Vogt, S.~Moch and J.~A.~M.~Vermaseren,
  ``The three-loop splitting functions in QCD: The singlet case,''
  Nucl.\ Phys.\  B {\bf 691}, 129 (2004)
  [arXiv:hep-ph/0404111].
}

\lref\MochTM{
  S.~Moch, J.~A.~M.~Vermaseren and A.~Vogt,
  ``Three-loop results for quark and gluon form factors,''
  Phys.\ Lett.\  B {\bf 625}, 245 (2005)
  [arXiv:hep-ph/0508055].
}

\lref\MochID{
  S.~Moch, J.~A.~M.~Vermaseren and A.~Vogt,
  ``The quark form factor at higher orders,''
  JHEP {\bf 0508}, 049 (2005)
  [arXiv:hep-ph/0507039].
}

\lref\KotikovER{
  A.~V.~Kotikov, L.~N.~Lipatov, A.~I.~Onishchenko and V.~N.~Velizhanin,
  Phys.\ Lett.\  B {\bf 595}, 521 (2004)
  [Erratum-ibid.\  B {\bf 632}, 754 (2006)]
  [arXiv:hep-th/0404092].
}

\lref\DrummondCF{
  J.~M.~Drummond, J.~Henn, G.~P.~Korchemsky and E.~Sokatchev,
  Nucl.\ Phys.\  B {\bf 795}, 52 (2008)
  [arXiv:0709.2368 [hep-th]].
}

\lref\BernIZ{
  Z.~Bern, L.~J.~Dixon and V.~A.~Smirnov,
  Phys.\ Rev.\  D {\bf 72}, 085001 (2005)
  [arXiv:hep-th/0505205].
}

\lref\KorchemskayaJE{
  I.~A.~Korchemskaya and G.~P.~Korchemsky,
  Phys.\ Lett.\  B {\bf 287}, 169 (1992).
}

\lref\AldayHR{
  L.~F.~Alday and J.~M.~Maldacena,
  JHEP {\bf 0706}, 064 (2007)
  [arXiv:0705.0303 [hep-th]].
}

\lref\AldayYW{
  L.~F.~Alday and R.~Roiban,
  Phys.\ Rept.\  {\bf 468}, 153 (2008)
  [arXiv:0807.1889 [hep-th]].
}

\lref\AldayHE{
  L.~F.~Alday and J.~Maldacena,
  JHEP {\bf 0711}, 068 (2007)
  [arXiv:0710.1060 [hep-th]].
}

\lref\McGreevyKT{
  J.~McGreevy and A.~Sever,
  JHEP {\bf 0802}, 015 (2008)
  [arXiv:0710.0393 [hep-th]].
}

\lref\DrummondAUA{
  J.~M.~Drummond, G.~P.~Korchemsky and E.~Sokatchev,
  Nucl.\ Phys.\  B {\bf 795}, 385 (2008)
  [arXiv:0707.0243 [hep-th]].
}

\lref\BeccariaTG{
  M.~Beccaria, V.~Forini, A.~Tirziu and A.~A.~Tseytlin,
  Nucl.\ Phys.\  B {\bf 812}, 144 (2009)
  [arXiv:0809.5234 [hep-th]].
}

\lref\FreyhultMY{
  L.~Freyhult and S.~Zieme,
  arXiv:0901.2749 [hep-th].
}

\lref\FioravantiXT{
  D.~Fioravanti, P.~Grinza and M.~Rossi,
  arXiv:0901.3161 [hep-th].
}

\lref\AldayMF{
  L.~F.~Alday and J.~M.~Maldacena,
  JHEP {\bf 0711}, 019 (2007)
  [arXiv:0708.0672 [hep-th]].
}

\lref\MuellerIH{
  A.~H.~Mueller,
  Phys.\ Rev.\  D {\bf 20}, 2037 (1979).
}

\lref\StermanQN{
  G.~Sterman and M.~E.~Tejeda-Yeomans,
  Phys.\ Lett.\  B {\bf 552}, 48 (2003)
  [arXiv:hep-ph/0210130].
}

\lref\BelitskyTC{
  A.~V.~Belitsky,
  Phys.\ Lett.\  B {\bf 442}, 307 (1998)
  [arXiv:hep-ph/9808389].
}

\lref\KorchemskySI{
  G.~P.~Korchemsky,
  Mod.\ Phys.\ Lett.\  A {\bf 4}, 1257 (1989).
}

\lref\KorchemskyXV{
  G.~P.~Korchemsky and G.~Marchesini,
  Nucl.\ Phys.\  B {\bf 406}, 225 (1993)
  [arXiv:hep-ph/9210281].
}

\lref\KRone{
  G.~P.~Korchemsky and A.~V.~Radyushkin,
  Phys.\ Lett.\  B {\bf 171}, 459 (1986).
}

\lref\KRtwo{
  S.~V.~Ivanov, G.~P.~Korchemsky and A.~V.~Radyushkin,
  Yad.\ Fiz.\  {\bf 44}, 230 (1986)
  [Sov.\ J.\ Nucl.\ Phys.\  {\bf 44}, 145 (1986)].
}

\lref\KRthree{
  G.~P.~Korchemsky and A.~V.~Radyushkin,
  Sov.\ J.\ Nucl.\ Phys.\  {\bf 45}, 127 (1987)
  [Yad.\ Fiz.\  {\bf 45}, 198 (1987)].
}

\lref\KRfour{
  G.~P.~Korchemsky and A.~V.~Radyushkin,
  Sov.\ J.\ Nucl.\ Phys.\  {\bf 45}, 910 (1987)
  [Yad.\ Fiz.\  {\bf 45}, 1466 (1987)].
}

\lref\RyangBC{
  S.~Ryang,
  Phys.\ Lett.\  B {\bf 659}, 894 (2008)
  [arXiv:0710.1673 [hep-th]].
}

\lref\RavindranMB{
  V.~Ravindran, J.~Smith and W.~L.~van Neerven,
  Nucl.\ Phys.\  B {\bf 704}, 332 (2005)
  [arXiv:hep-ph/0408315].
}

\lref\IdilbiDG{
  A.~Idilbi, X.~d.~Ji and F.~Yuan,
  Nucl.\ Phys.\  B {\bf 753}, 42 (2006)
  [arXiv:hep-ph/0605068].
}

\lref\BecherMR{
  T.~Becher, M.~Neubert and B.~D.~Pecjak,
  JHEP {\bf 0701}, 076 (2007)
  [arXiv:hep-ph/0607228].
}

\lref\BecherTY{
  T.~Becher, M.~Neubert and G.~Xu,
  JHEP {\bf 0807}, 030 (2008)
  [arXiv:0710.0680 [hep-ph]].
}

\lref\CollinsBT{
  J.~C.~Collins,
  Adv.\ Ser.\ Direct.\ High Energy Phys.\  {\bf 5}, 573 (1989)
  [arXiv:hep-ph/0312336].
}

\lref\MaldacenaRE{
  J.~M.~Maldacena,
  Adv.\ Theor.\ Math.\ Phys.\  {\bf 2}, 231 (1998)
  [Int.\ J.\ Theor.\ Phys.\  {\bf 38}, 1113 (1999)]
  [arXiv:hep-th/9711200].
}

 \lref\gkp{
  S.~S.~Gubser, I.~R.~Klebanov and A.~M.~Polyakov,
  ``A semi-classical limit of the gauge/string correspondence,''
  Nucl.\ Phys.\  B {\bf 636}, 99 (2002)
  [arXiv:hep-th/0204051].
}

\lref\KotikovAB{
  A.~V.~Kotikov and L.~N.~Lipatov,
  Nucl.\ Phys.\  B {\bf 661}, 19 (2003)
  [Erratum-ibid.\  B {\bf 685}, 405 (2004)]
  [arXiv:hep-ph/0208220].
}

\lref\AldayCG{
  L.~F.~Alday,
  Fortsch.\ Phys.\  {\bf 56}, 816 (2008)
  [arXiv:0804.0951 [hep-th]].
}

\lref\KorchemskyUZ{
  G.~P.~Korchemsky and G.~Marchesini,
  Phys.\ Lett.\  B {\bf 313}, 433 (1993).
}


\Title{\vbox{\baselineskip12pt \hbox{} \hbox{
} }} {\vbox{\centerline{Universal structure of subleading infrared poles
  }
\centerline{at strong coupling
}
}}
\bigskip
\centerline{Luis F. Alday}
\bigskip
\centerline{\it School of Natural Sciences, Institute for
Advanced Study} \centerline{\it Princeton, NJ 08540, USA}

\vskip .3in \noindent
Recently a concise expression for the subleading infrared singularity of dimensional-regularized gauge theories has been proposed. For conformal theories,
such relation involves a universal eikonal contribution plus a non-eikonal contribution, related to the subleading term in the anomalous dimension of twist two operators with large spin. In this note we make use of the $AdS/CFT$ correspondence in order to check such conjecture at strong coupling for the case of ${\cal N}=4~ SYM$.

\Date{ }

\newsec{Introduction}

On-shell scattering amplitudes are perhaps the most basic quantities computed in any gauge theory. A common feature of on-shell scattering amplitudes in massless gauge theories in four dimensions is the presence of infra-red (IR) divergences, originating from low energy virtual particles as well as from virtual momenta almost parallel to the external ones. In order to define the amplitudes properly, an IR regulator should be introduced. A convenient choice is the use of (a version of) dimensional regularization, in which we consider the theory in $D=4-2\epsilon$ dimensions. IR divergences, then, manifest themselves as poles of the form ${1 \over \epsilon}$. Such IR divergences have a structure that is captured by the soft/collinear factorization theorem, see {\it e.g.} \refs{\MuellerIH,\CollinsBT,\StermanQN} , of the schematic form \foot{When writing this expression we have in mind conformal theories. Furthermore, strictly speaking, the functions that appear in this expression are $f^{-2}(\lambda)$ and $g^{-1}(\lambda)$, with some extra proportionality factors.}
\eqn\expon{{\cal A}_{div} \approx e^{{1 \over \epsilon^2}f(\lambda)+{1 \over \epsilon}G(\lambda)}}
where $f(\lambda)$ is the so called cusp anomalous dimension \refs{\KRone,\KRtwo,\KRthree,\KRfour} and $G(\lambda)$ is the so called collinear anomalous dimension. The cusp anomalous dimension appears in several other computations. In particular, it controls the anomalous dimension of twist two operators, of the form $Tr \phi D^S \phi$, in the large spin limit \refs{\KorchemskySI,\KorchemskyXV}, namely
\eqn\twisttwo{\Delta-S=f(\lambda)\log S-B(\lambda)+...}
The cusp anomalous dimension is a very well known quantity. On the other hand, much less is known about the "subleading" quantities $G(\lambda)$ and $B(\lambda)$. In particular, they are not universal, in the sense that they depend on the kind of particles under consideration. Recently, Dixon, Magnea and Sterman (DMS), based on previous hints from \refs{\RavindranMB,\MochTM,\IdilbiDG,\BecherMR,\BecherTY} , put forward an interesting proposal \DixonGR\ , according to which the difference $G(\lambda)-2B(\lambda)$ is a universal quantity and can be obtained from a eikonal contribution\foot{In the eikonal approximation hard partonic lines are replaced by Wilson lines.} $G_{eik}$ to the collinear anomalous dimension ( plus a term proportional to the beta function that will vanish for the case considered in this note).

An interesting theory in which this set of ideas can be tested is maximally super-symmetric Yang-Mills (MSYM). Due to its high degree of symmetries, this theory is much simpler to study perturbatively than, for instance, $QCD$. On the other hand, the strong coupling limit of the theory can be studied by means of the $AdS/CFT$ duality \MaldacenaRE\ , by studying a weakly coupled sigma model.

The aim of this note is to study the universality proposed in \DixonGR\ at strong coupling for the case of MSYM. In the first half of the note we review the relation at weak coupling. It turns out that at two loops the above mentioned eikonal contribution can be extracted from the expectation value of a "rectangular" light-like Wilson loop. In the second half of the note, the relation is studied at strong coupling, where universality is also observed. Furthermore, we will see that the eikonal contribution can be again extracted from the expectation value of the rectangular light-like Wilson loop. Since the computation of $G$ at strong coupling is also formally equivalent to a Wilson loop computation \refs{\AldayHR,\AldayYW}, naively one would say that $G$ and $G_{eik}$ coincide at strong coupling. However, $G_{eik}$ should be computed in the original $AdS$ background as opposed to the $T-$dual $AdS$ background. While the "unregularized" spaces are equivalent, they differ when we use dimensional regularization and according to the DMS relation, $B_{\delta}$ measures this difference. Finally we end up with some conclusions. The relevant computations at strong coupling are deferred to the appendices.

\newsec{Dixon-Magnea-Sterman relation at weak coupling}

Dixon, Magnea and Sterman studied in \DixonGR\ the subleading soft and collinear poles of form factors and amplitudes in dimensionally-regulated massless gauge theories. This poles are characterized by a function $G(\alpha_s)$, which in general depends on both, the spin and gauge quantum numbers of the particles under consideration. For the case of conformal theories, DMS wrote this function as the sum of two contributions: a universal eikonal anomalous dimension \foot{More precisely, $G_{eik}$ carries no information about the spin of the parton, only is representation under the gauge group.} $G_{eik}(\alpha_s)$ and a non eikonal contribution $B_{\delta}(\alpha_s)$, given by the virtual contribution to the Altarelli-Parisi splitting kernel. The proposed relation for the particular case of ${\cal N}=4$ SYM reads \foot{A similar relation, apparently different at two loops, appeared in \KorchemskySI\ , eq. (33).}

\eqn\univ{G(\alpha_s)=G_{eik}(\alpha_s)+2B_{\delta}(\alpha_s)}
In the following we study in detail each of these terms at the perturbative level and work out some examples of the above relation

\subsec{$G(\alpha_s)$}

$G(\alpha_s)$ is given by the subleading IR pole in the Sudakov form factor. The perturbative result for the case of gluons, up to three loops, has been presented in  \MochTM\  , eq. (18). According to the maximally transcendentality principle \KotikovAB\ , we expect the leading transcendental piece to correspond to the MSYM result

\eqn\ggluon{G_g^{(2)}=-4 C_A^2 \zeta_3,~~~~~G_g^{(3)}=8 C_A^3 ({10
\over 3 } \zeta_2 \zeta_3+4 \zeta_5)}
from now on $C_{A,F}$ denote the Casimirs of the adjoint/fundamental representation. For most of the discussion in theses notes, we will restrict ourselves to the planar limit, where $C_A=2C_F=N$.
The loop quantities defined above are the coefficient of
$\left({\alpha_s \over 4 \pi }\right)^L$. On the other hand, the coupling $a$ used in \BernIZ\ is
$a={\alpha_s N \over
2\pi}$. After taking this into account we find perfect agreement between $G$ used here and what was called $\hat
{\cal G}_0$ in \BernIZ\ .

The quark form factors can be found in eq. (3.10) of \MochID\ . Extracting the pieces with leading transcendentality we obtain

\eqn\gquark{\eqalign{G_q^{(2)}&=48 C_F^2 \zeta_3-52 C_F C_A
\zeta_3\cr G_q^{(3)}&=-32 C_F^3(2 \zeta_2 \zeta_3+15 \zeta_5)+16
C_F^2 C_A(2 \zeta_2 \zeta_3+15\zeta_5)+C_F C_A^2({176 \over
3}\zeta_2 \zeta_3+272 \zeta_5) }}

\subsec{ $B_\delta(\alpha_s)$}

In eq. \univ\ , $B_{\delta}(\alpha_s)$  is the coefficient of
$\delta(1 - x)$ in the Altarelli-Parisi diagonal splitting
function.

\eqn\ap{P_{ii}(x)={\gamma_K^{[i]}(\alpha_s) \over 2} \left[{1
\over 1-x} \right]_+ + B_{\delta}^{[i]}(\alpha_s)\delta(x-1)+...}
Which is related to the large spin behavior of twist two
operators, see for instance \BernIZ\

\eqn\leadingj{\gamma(S)={1 \over 2}\gamma_k(\alpha_s)(\ln
S+\gamma_e)-B(\alpha_s)+...}
At two and
three loops, $P_{qq}$ and $P_{gg}$, where $g$ denotes a gluon and $q$ a quark, have been computed in \MochPA\
and \VogtMW\ . We can extract the higher transcendentality terms
proportional to $\delta(1-x)$

\eqn\B{\eqalign{ B_{gg}^{(2)}&=12 C_A^2
\zeta_3,~~~~~B_{gg}^{(3)}=-16C_A^3(\zeta_2 \zeta_3+5 \zeta_5) \cr
B_{qq}^{(2)}&=-12 C_A C_F \zeta_3+24 C_F^2 \zeta_3,\cr
B_{qq}^{(3)}&=16C_A C_F^2(\zeta_2\zeta_3+{15 \over 2 }\zeta_5)+40
C_A^2 C_F \zeta_5-16C_F^3(2\zeta_2 \zeta_3+15\zeta_5) }}

\subsec{Universal quantity $G_{eik}(\alpha_s)$}

From the above two and three loops results we can consider the difference $G-2B$ for the different cases

\eqn\univtt{ \eqalign{G_g^{(2)}-2 B_{gg}^{(2)}&= -28 C_A^2
\zeta_3,~~~~~G_q^{(2)}-2 B_{qq}^{(2)}= -28 C_A C_F \zeta_3 \cr
G_g^{(3)}-2 B_{gg}^{(3)}&= {8 \over 9} C_A^3(11 \pi^2 \zeta_3+216
\zeta_5 ),~~~~~G_q^{(3)}-2 B_{qq}^{(3)}= {8 \over 9} C_A^2 C_F(11
\pi^2 \zeta_3+216 \zeta_5 )} }
In agreement with the claim of \DixonGR\ , $G-2B$ is a universal
quantity, up to a factor depending on the color representation (in this case $C_A$ vs. $C_F$). This difference is exactly what was called $f$ \foot{Not to be confused with the cusp anomalous dimension.} in \MochTM\ and, as already
noted there, is a universal quantity.

The following remark will be important for what will be discussed in these notes. According to \DixonGR\ , the eikonal contribution $G_{eik}$ is related to the function responsible for soft single logarithms in threshold resummation for the Drell-Yan process. Indeed, it can be extracted, at two loops, from the Drell-Yan anomalous dimension computed in \BelitskyTC\ . On the other hand, as also mentioned in \DixonGR\ , note that at two loops,
$G_{eik}^{(2)}=-28 C_A^2 \zeta_3$, coincides with the two loops subleading pole of the expectation value of the rectangular light-like Wilson loop considered in \KorchemskayaJE\ and \DrummondCF\ . In particular, $G_{eik}$ (which was called $\Gamma(a)$ in \DrummondCF\ ) can be extracted from the renormalization group equation for the expectation value of such Wilson loop, of the form
\eqn\rgetwo{{\partial \log \langle W \rangle \over \partial \log \mu^2}=- {f(a) \over 2} \log x_{13}^2 x_{24}^2 \mu^4 -G_{eik}(a)-{1 \over \epsilon} \int_0^a {da' \over a'}f(a') +{\cal O}(\epsilon)}
where $f(a)$ is the cusp anomalous dimension and the $x_i's$ denote the position of the cusps. From eq. (30) of \DrummondCF\ we get the desired result (after taking into account the difference of conventions). In the next section we will assume that $G_{eik}$ can be extracted from the same computation at strong coupling and we will see that indeed the DMS relation continues to hold.

The connection between the Drell-Yan and rectangular light-like Wilson loop computations can probably be understood as a consequence of conformal invariance. The Drell-Yan computation corresponds to two single cusp Wilson loops, whose cusps are separated by a time-like distance \KorchemskyUZ\ . It can be seen, {\it e.g.} in \RyangBC\ , that there exist conformal transformations taking the rectangular Wilson loop world-sheet to the kind of world-sheets relevant to Drell-Yan processes. Such world-sheets, however, are hard to analyze, since the radial coordinate is in general complex and they do not seem to be Euclidean everywhere.

\newsec{Connection to strong coupling}

If the relation \univ\ is to hold to all orders in perturbation theory, one may expect that there is some way to check it at strong coupling, by using the $AdS/CFT$ duality. Provided we make the assumption that $G_{eik}$ can be extracted from the rectangular light-like Wilson loop, all the ingredients in this relation can be easily computed (or have been already computed!) at strong coupling, as shown in appendix A.

Scattering amplitudes of MSYM can be computed at strong coupling by using the $AdS/CFT$ duality \AldayHR . Very much as in the gauge theory, a regulator needs to be introduced in order to define the amplitudes properly. In \AldayHR\  the analogous of dimensional regularization was used, which allowed to compute the strong coupling limit of the functions characterizing the IR poles of the amplitudes . The collinear anomalous dimension $G$ was computed at strong coupling for gluons in \AldayHR\ and quarks in \refs{\AldayHE,\McGreevyKT} .

\eqn\gs{\eqalign{G_{gluon}={\sqrt{\lambda} \over 2 \pi}(1-\log
2)\cr G_{quark}={\sqrt{\lambda} \over 4 \pi}(1-3\log 2)}}
We have divided by 2 what was called $g_{quark}$ in \AldayHE\ , since this is the quantity that enters in the form factor.

The function $B$ at strong coupling can be computed by considering classical strings spinning on $AdS$. According to the dictionary of the $AdS/CFT$ correspondence, such strings states corresponds to twist two operators with high spin \gkp\  and the energy of the former is related to the anomalous dimension of the later. A detailed computation is shown in the appendix, the final result for gluons and quarks is\foot{Quarks, transforming in the fundamental representation, can be considered in ${\cal N}=4$ SYM by adding a flavor symmetry.}

\eqn\bs{\eqalign{B_{gg}={\sqrt{\lambda} \over 2\pi}(\ln
\left({\sqrt{\lambda} \over 2\pi} \right)+1 -2\log 2 +\gamma_e ) \cr
B_{qq}={\sqrt{\lambda} \over 4 \pi}(\ln \left({\sqrt{\lambda} \over
2\pi} \right)+1 -3\log 2 +\gamma_e) }}
In order to extract the functions $B$ from the computation in the appendix one needs to use the precise relation \leadingj . On the other hand, as explained in appendix B, one needs to divide by an extra factor of 2, coming from the use of different conventions in the computations at weak and strong coupling.

From this results we see that $G_R-2B_R=f_R X$, with X some
universal function and $f_R$ a factor that depends on the representation ({\it e.g.} $f_{R=q}=1$ for quarks and  $f_{R=g}=2$ for gluons). For instance, focusing on the case of the gluons we obtain

\eqn\gluondif{
G_{gluon}-2B_{gg}={ \sqrt{\lambda} \over 2\pi}(-1-2 \gamma_e +5 \log{2}+2 \log{\pi}-\log{\lambda})
}
This difference should be then compared to $G_{eik}$ at strong coupling, extracted from the rectangular Wilson loop computation done in the appendix

\eqn\Geikstrong{G_{eik}={\sqrt{\lambda} \over 2\pi}( -1-2\gamma_e+5\log 2+2 \log \pi-\log \lambda)}
Hence we see that the DMS relation holds at strong coupling. Several comments are in order. First, note that the strong coupling computations giving $G$ and $G_{eik}$ are very similar, with the difference that $G$ is computed in the dual coordinates and $G_{eik}$ is computed in the original coordinates, since the former comes from a scattering computation while the later comes from a Wilson loops computation. The two computations are naively the same, however dimensional regularization acts on a different way. We could interpret the DMS relation as telling us what the difference $G-G_{eik}$ should be. Second, note that the precise matching depends on several factors having to do with conventions, etc, and as such is not very robust. On the other hand, the matching of the term $-1-\log \lambda$ is more neat and still non trivial. Finally, we have assumed that it is possible to extract $G_{eik}$ from the rectangular Wilson loop computation, which is true at two loops and, according to the results of this note, also at strong coupling.

Irrespective of the strong coupling result for $G_{eik}$, a universality for the combination $G-2B$ is observed at strong coupling.  This universality seems to persist even when using other schemes, such as the radial cut-off introduced in \AldayCG\ . Furthermore, note that most of our strong coupling computations are not restricted to four dimensions , so universality can extend to other dimensions as well. It would be interesting to study these issues in more detail.

\newsec{Conclusions}

In this note we have tested the universality relation \univ , between the subleading pieces of several computations, at strong coupling, by using the $AdS/CFT$ duality. One can explicitly compute $B_\delta(\lambda)$ and $G(\lambda)$ at strong coupling for quarks and gluons and check that indeed universality holds (up to a factor of two, which exactly coincides with the ratio $C_A/C_F$ in the planar limit)

Besides, we have assumed that $G_{eik}$ can be extracted from the dimensionally-regularized rectangular light-like Wilson loop. One can explicitly check that this is the case at two loops and the results of this note imply that this is the case at strong coupling too. This way of computing $G_{eik}$ does not follow immediately from the definition of \DixonGR\ . According to  \DixonGR\ , $G_{eik}$ can be extracted from the Drell-Yan anomalous dimension, computed for instance at two loops in \BelitskyTC . The equivalence of both computations is possibly a consequence of conformal symmetry, however we have not proven this statement.

The computation presented here is a non trivial check of the relation \univ\ and hopefully it will help in order to shed some light in the understanding of collinear anomalous dimensions in scattering amplitudes of massless gauge theories. Note that in particular, \univ\ implies that the difference $G-G_{eik}$ can be written in terms of the anomalous dimension of an operator. Actually, an integral equation can be written for $B_\delta$ \foot{At least without taking into account wrapping effects. Whether such effects arise is not clear.}, which allows to compute it for any value of the coupling constant \refs{\FreyhultMY,\FioravantiXT}.

Since the relation \univ\ is expected to be true for any value of the coupling constant, one may expect that it is a consequence of symmetries and maybe can be proven along the lines of \AldayMF\ . Furthermore, if such relation is a consequence of symmetries, a version of it may hold beyond the case of dimensionally regulated four dimensional theories.

{ \bf Acknowledgments }

We would like to thank J. Maldacena for collaboration at early stages of this project. We would like to thank L. Dixon, G. Korchemsky, L. Magnea and J. Maldacena for illuminating discussions and comments on the draft. This work   was  supported in part by U.S.~Department of Energy
grant \#DE-FG02-90ER40542.

\appendix{A}{Strong coupling coumputations of $B_\delta$ and $G_{eik}$}

\subsec{$B_\delta$ at strong coupling}

The anomalous dimension of high spin operators at strong coupling can be studied by considering classical spinning strings, as in\gkp\ , to which we refer the reader for the details of the following computation.

We would like to compute the constant $b(\lambda)$ in the large spin expansion of the anomalous dimension \eqn\constb{ \Delta -
S = f(\lambda) \log S + b(\lambda) + {\cal O}( \log S/S) }
In formulas (17) and (18) of \gkp\ an expression for
the energy and the spin is given in terms of a parameter $\rho_0$, and in particular

\eqn\energmis{
\Delta - S = 2 { R^2 \over  \pi \alpha'}
 \int_0^{\rho_0} d\rho { \sinh \rho_0 - e^{-\rho_0 } \sinh^2 \rho \over
 \sqrt{ \sinh^2 \rho_0 - \sinh^2 \rho}  } \equiv 2 { R^2 \over  \pi \alpha'} I,~~~~~~~~{R^2 \over \alpha' } = \sqrt{\lambda}
}
 We also see from their formulas (29) and (31) that
 \eqn\vals{
 S =  { R^2 \over 2 \pi \alpha'}  e^{2 \rho_0}+...~~~\rightarrow~~~2 \rho_0 = \log S - \log ({ R^2
\over 2 \pi \alpha'})+...
}
We now evaluate the integral in \energmis\ which we write as
\eqn\integra{\eqalign{ I=\rho_0 + \int_0^{\rho_0} d\rho { \sinh
\rho_0  - \sqrt{ \sinh^2 \rho_0 - \sinh^2 \rho} - e^{- \rho_0 }
\sinh^2 \rho \over \sqrt{ \sinh^2 \rho_0 - \sinh^2 \rho}  } }}
Now
we take the $\rho_0 \to \infty$ in this last integral. Naively we
would say that the result is zero. On the other hand, there is an
end point contribution near $\rho \sim \rho_0$ which can be computed as follows

 \eqn\endpointc{\eqalign{
 & \int_0^{\rho_0} d\rho { \sinh \rho_0  - \sqrt{ \sinh^2 \rho_0 - \sinh^2 \rho} - e^{- \rho_0 } \sinh^2 \rho
\over \sqrt{ \sinh^2 \rho_0 - \sinh^2 \rho}  } \to \cr & \to
\int_{-\infty}^0 dx { 1 - \sqrt{ 1 - e^{2 x} } - {1 \over 2 } e^{
2 x} \over \sqrt{ 1 - e^{2 x} } } = -1/2 + \log 2 }} where $x =
\rho - \rho_0 $. Putting these results together we find that
\eqn\findre{ \Delta - S = { \sqrt{\lambda} \over \pi } \left[ 2 \rho_0 +   (
-1 + 2 \log 2 )  \right] }
Then we get our final result \eqn\finalres{ b =
 { \sqrt{\lambda} \over \pi} \left[ - \log ({ \sqrt{\lambda} \over 2 \pi } ) + (-1 + 2 \log 2 ) \right]
}
This agrees exactly with the result obtained for instance in \BeccariaTG . From this result, we can easily extract the strong coupling limit of $B_{gg}(\lambda)$.

The same computation can be easily repeated for an open string, which is
related to an operator of the form $ \bar q D^S q$. The
energy and spin are given by the same integrals as before except
that we divide the right hand sides by a factor of two. Thus, as a
function of $\rho_0$, $\Delta -S$ is half of what it was before.
Thus we find \eqn\finalresqq{\eqalign{ & \Delta - S |_{\bar q q }
= {  \sqrt{\lambda} \over 2 \pi } \left[ 2 \rho_0 +   ( -1 + 2 \log 2 )
\right] = { f \over 2} \log S + { b_{\bar q q } \over 2 } \cr &
b_{\bar q q} = {  \sqrt{\lambda} \over   \pi} \left[ - \log ({  \sqrt{\lambda} \over 2 \pi
} ) + (-1 + 2 \log 2 ) + \log 2  \right]=
 { \sqrt{\lambda} \over  \pi} \left[ - \log ({ \sqrt{\lambda} \over 2 \pi } ) -1 + 3 \log 2 \right]
}} The extra $\log 2$ comes from the fact that now \eqn\valsqq{
 S =  { 1 \over 2 } { R^2 \over 2 \pi \alpha'}  e^{2 \rho_0}
}
instead of the equation \valsqq . From this computation we can extract the strong coupling limit of what we would like to call $B_{q\bar q}(\lambda)$. As a final remark, note that the way $B_{qq}$ is computed from the corresponding computation for gluons (diving by two in the right places) is very much the same as the way $g_{quark}$ was computed from the corresponding computation for gluons in \AldayHE\ . This is part of the reason for universality at strong coupling.

\subsec{$G_{eik}$ at strong coupling}

According to the discussion in the body of these notes, the universal factor $G_{eik}$ should be given by the collinear anomalous piece of the rectangular Wilson loop. Note that since we are computing a Wilson loop expectation value, we have to use the original coordinates (as opposed to the T-dual coordinates).  The dimensional regularized metric is

\eqn\gravdual{\eqalign{ ds^2 = &  f^{-1/2} dx_{D}^2 + f^{1/2}
[dr^2  + r^2 d\Omega^2_{9-D}]~,~~~~~~~~~~~~~~D = 4 - 2 \epsilon
\cr f = & { c_D \lambda \over r^{ 8-D} } ~,~~~~~~~~~~~~~~~c_{D} =
2^{4 \epsilon } \pi^{ 3\epsilon} \Gamma( 2 + \epsilon )
{{\mu^{2\epsilon}}\over {(4\pi e^{-\gamma})^\epsilon}}}}
We can compute the Nambu-Goto action in the usual way. The Log of the Wilson loop at strong coupling will then be directly related to the Nambu-Goto area. Let us first discuss the single cusp solution, which can be embedded into $AdS_3$

\eqn\ampstrong{ A={1 \over 2\pi} \int \sqrt{(\partial_x r)^2-(\partial_t r)^2-{1 \over f}}}
 The solution corresponding to the single cusp can be found for
arbitrary values of $\epsilon$.

\eqn\singlecuspor{ r(x,t)={k_1 \over
(t^2-x^2)^{k_2}},~~~~~k_1=\left({2+3 \epsilon+\epsilon^2 \over c_D
\lambda} \right)^{-1 \over 2+2\epsilon} ,~~~~~k_2={1 \over 2+2
\epsilon} }
Note that the square root behavior of the $t^2-x^2$ dependence
is modified by $\epsilon$. Also, note that the boundary is located at $r =\infty$. Next, we focus on the four edges solution. It is convenient to embed the surface in Poincare coordinates $(r,t,x_1,x_2)$ and parametrize the world-sheet by its projection to the $(x_1,x_2)$ plane.

\eqn\ampstrong{ A={1 \over 2\pi} \int {1 \over \sqrt{c_D \lambda}}\sqrt{c_D \lambda((\partial_i r)^2- (\partial_1 t \partial_2 r-\partial_1 r \partial_2 t)^2) -r^{4+2\epsilon}(-1+(\partial_i t)^2)}}
Choosing an approximate solution with the correct behavior close to the cusps (and solving the equations of motion for $\epsilon=0$) we obtain

\eqn\fouredges{ r(x_1,x_2)= k_0 \left({c_D \lambda \over (1-x_1^2)(1-x_2^2)} \right)^{1 \over 2+2\epsilon},~~~t(x_1,x_2)=x_1 x_2,~~~~~k_0=1+{\cal O}(\epsilon) }
subleading terms in $k_0$ will drop out from our final result. Plugging this solution into the Nambu-Goto action and using the evolution equation we obtain

\eqn\evol{{\partial A \over \partial \log \mu^2}={\sqrt{\lambda} \over \pi \epsilon}+\sqrt{\lambda} {\log 8 \mu^2 -1+\log 8+2\log \pi -\log \lambda \over 2\pi}+{\cal O}(\epsilon) }
Note that the presence of the factor $\log \lambda$ is a direct consequence of the fact that the solution \fouredges\ depends explicitly on $\lambda$ and this dependence is modified by the presence of $\epsilon$ . This is a very important difference between original and $T-$dual variables.The expectation value of light-like Wilson loops is ultra-violet (UV) divergent and as such the scale $\mu$ appearing in \evol\ is a UV scale. One can see from \DrummondAUA\ , eqs. (15) and (44) that the UV and IR cut-offs are of the form $\mu^2_{IR}=4\pi e^{-\gamma_e} \tilde{\mu}^2$ and $\mu^{-2}_{UV}=\pi e^{\gamma_e} \tilde{\mu}^2$, hence $\mu^2_{IR} \mu^2_{UV}=4 e^{-2\gamma_e}$. When comparing $G_{eik}$ to IR divergent quantities (such as scattering amplitudes), we need to rewrite $\mu_{UV}$ in terms of $\mu_{IR}$. This results in a extra shift for $G_{eik}$, the final result being

\eqn\Geikstronga{G_{eik}={\sqrt{\lambda} \over 2\pi}( -1-2\gamma_e+2 \log \pi+5\log 2-\log \lambda)}
Note that we have used the fact that $x_{i,i+2}^2=8$ for the rectangular Wilson loop under consideration.

\appendix{B}{Understanding the factor of two}

The factor of two in the definition of $B$ between weak and strong coupling computations has a simple
explanation due to the use of different conventions. It is
instructive to look at the paper \KotikovER\ . There, the authors compute
the anomalous dimension of twist two operators in ${\cal N}=4$ SYM
up to terms of order $(\log{S})^0$. They write their result as
follows

\eqn\weakexp{
\gamma(S)=\hat{a}\gamma_{uni}^{(0)}+\hat{a}^2\gamma_{uni}^{(1)}+\hat{a}^3\gamma_{uni}^{(2)}+...,
~~~~~\hat{a}={\alpha N_c \over 4 \pi} }
The large $S$ expansions of the above terms are also given in that
paper, eqns. (26)-(28)

\eqn\largej{\eqalign{
\gamma_{uni}^{(0)}(S)=-4(\ln{S}+\gamma_e)+0+...\cr
\gamma_{uni}^{(1)}(S)=8 \zeta_2(\ln{S}+\gamma_e)+12 \zeta_3 +...
 \cr \gamma_{uni}^{(2)}(S)=-88
\zeta_4(\ln{S}+\gamma_e)-16 \zeta_2 \zeta_3-80 \zeta_5+...}}
As the authors of \KotikovER\ mention, there is a difference of a factor of $-1/2$
between their conventions and the conventions of
Gubser-Klebanov-Polyakov (GKP) when they compute the cusp anomalous
dimension \gkp\ . We are also using GKP conventions, since we use their
calculation in order to compute the $B$ term at strong coupling. More precisely,
$\gamma_{LIP}=-1/2 \gamma_{GKP}$. Hence, if we translate what we
called $B$ at weak coupling to the GKP conventions, we obtain

\eqn\bweak{B_{gg}=24 \zeta_3 g^4-32(\zeta_2
\zeta_3+5\zeta_5)g^6+...}
where $g^2=\hat{a}={\alpha N_c \over 4\pi}$. Comparing this with
\B , we see that we get exactly twice the result. Hence, we need to divide by two the strong coupling result in order to adjust to the conventions usually used in perturbative computations and in particular used by Dixon, Magnea and Sterman in \DixonGR\ .

\listrefs

\bye